\documentclass{PoS}

\usepackage{url}

\title{Highlights of the MAGIC AGN program}

\ShortTitle{Highlights of the MAGIC AGN program}

\author{\speaker{Julian Sitarek}\\ 
        University of \L\'od\'z, PL-90236 Lodz, Poland\\
        E-mail: \email{jsitarek@uni.lodz.pl}}
\author{for the MAGIC Collaboration}

\abstract{
MAGIC is a system of two Cherenkov telescopes designed to perform observations of gamma rays with energies from about 50 GeV to tens of TeV. A low energy threshold and an excellent low energy performance make it a powerful instrument for studies of high energy processes in Active Galactic Nuclei (AGN). The MAGIC Collaboration dedicates about 40\% of the telescopes' time to such observations, both monitoring of known emitters, and Target of Opportunity observations of flaring events. In this presentation we will discuss the recent highlights of the MAGIC AGN program. In particular, multi year monitoring of radio galaxies, such as M87, and strong peculiar blazars, such as PG 1553+113, will be presented together with new discoveries based on the Target Of Opportunity program.
}

\FullConference{35th International Cosmic Ray Conference --- ICRC2017-\\
		10--20 July, 2017\\
		Bexco, Busan, Korea}

\begin{document}

\section{Introduction}
AGN (Active Galactic Nuclei) are the central parts of some of the galaxies, from which broadband, mainly non-thermal and highly variable radiation is observed.
They are believed to be powered by accretion of matter on a rotating black hole, which is also causing in some subclasses of those objects occurrence of superrelativistic outflows (the so-called jets) from their central engine. 
The broadband emission produced in AGN jets consists of two broad peaks; one in IR to X-ray range, and the second one in GeV-TeV $\gamma$ rays. 
While the first peak is commonly attributed to the synchrotron radiation, the origin of the second peak is still being disputed. 
In leptonic models the second peak is produced by inverse Compton scattering by electrons of ambient radiation fields, either internal to the jet (e.g. synchrotron-self-Compton model, see e.g. \cite{re67}), or external to it (see e.g. \cite{ds93}). 
AGN are observationally classified into multiple types. 
The most common in VHE (very-high-energy, $\gtrsim 100$\,GeV) $\gamma$-ray astronomy being BL Lacs, in which the jet direction is aligned with the line of sight to the observer causing strong enhancement of the observed flux by relativistic boosting. 
They are further divided into HBL, IBL and LBL (high, intermediate and low energy peaked BL Lacs respectively) sub-classes depending on the location of the SED peaks. 
Flat spectrum radio quasars (FSRQ), also strongly relativistically boosted, are intrinsically more powerful than BL Lacs, however their peaks are even more shifted to the lower energies, making their detection at VHE $\gamma$ rays challenging. 
The radiation fields external to the jet in FSRQ are also stronger than in BL Lacs, making those sources both more interesting and much more complicated. 
FSRQ and BL Lacs together are called blazars. 
Relativistically non-boosted counterparts of blazars are believed to be radio galaxies in which the jet orientation has a large angle to our viewing direction. 
The VHE $\gamma$-ray observations of AGN are giving us unique information about the highest energy processes occurring in them, and thus are crucial for understanding the extreme physical conditions in the jet of AGN.

MAGIC (\emph{Major Atmospheric Gamma Imaging Cherenkov}) is a system of two 17m diameter Imaging Atmospheric Cherenkov Telescopes \cite{al16a}. 
The telescopes are located at the height of 2200 m a.s.l. at the observatory of the Roque de los Muchachos in the Canary Island of La Palma ($28^\circ$N, $18^\circ$W). 
The first MAGIC telescope (MAGIC-I) performs observations since 2004. 
Starting from 2009, when the construction of the second telescope (MAGIC-II) finished, both telescopes are working together in the so-called stereoscopic mode. 
Large mirror area, combined with the stereo observations, provides unprecedented performance of the instrument at the lowest energies, namely around and below 100\,GeV \cite{al16b}. 
The low energy threshold ($\sim 50-60$~GeV at low zenith angle observations), combined with Northern hemisphere location (exploiting ample multiwavelength coverage of Northern sources) makes it an excellent instrument for observations of the VHE $\gamma$ rays from AGN objects. 
In this contribution we present the AGN observation program of MAGIC and discuss a few selected recent results.

\section{MAGIC AGN program}
\begin{figure}
\centering
\includegraphics[width=0.6\textwidth]{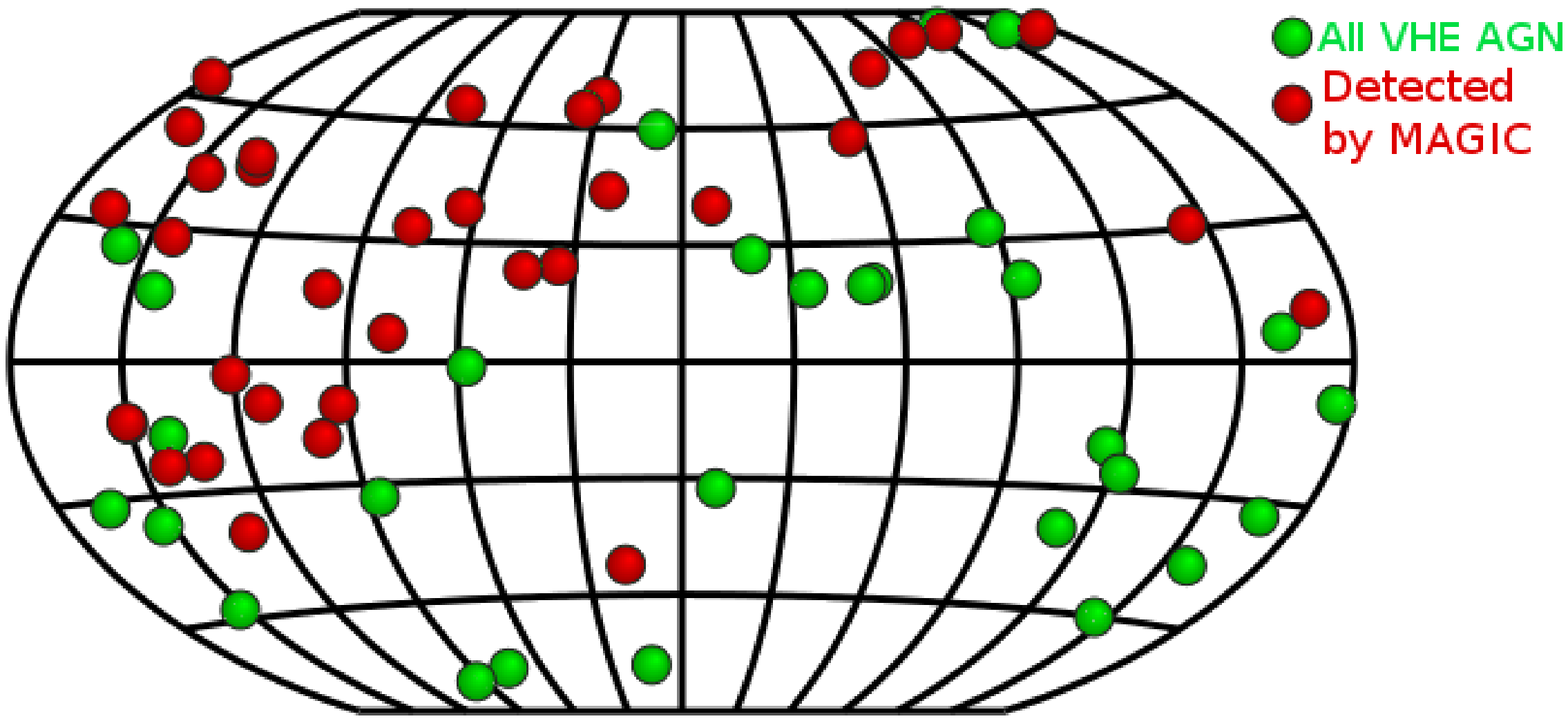}
\includegraphics[width=0.39\textwidth]{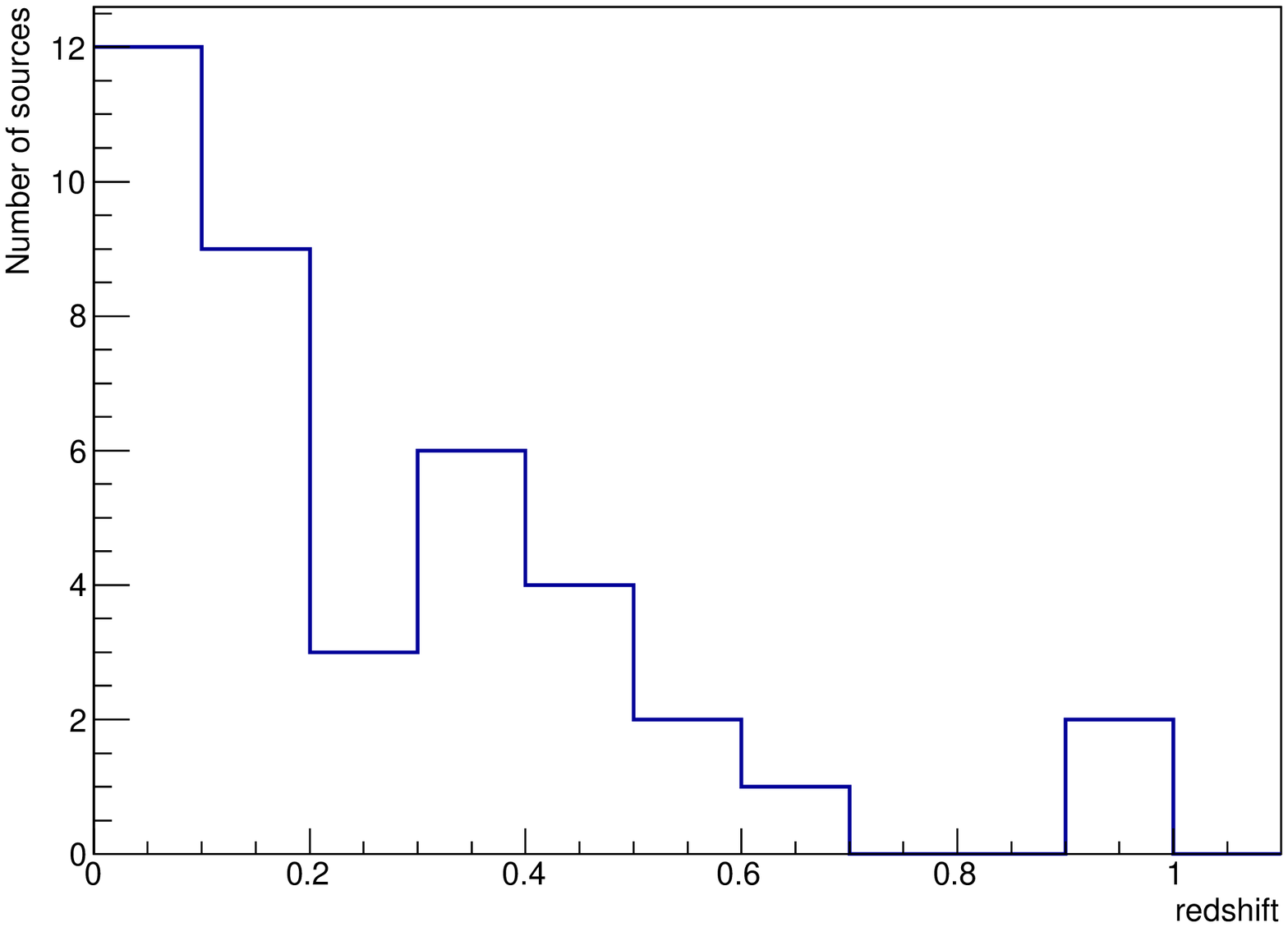}
\caption{
Left panel: extragalactic VHE $\gamma$-ray sky.
Green circles show all the currently detected in VHE $\gamma$ rays AGN, while those from which VHE $\gamma$-ray has been observed with MAGIC are shown with red circles. 
Figure created using maps from \cite{tevcat}. 
Right panel: the redshift distribution of AGN detected with MAGIC. 
}\label{fig:map}
\end{figure}

Over the last 25 years the VHE $\gamma$-ray astronomy of AGN objects evolved from a single known source to a moderately large catalog of 70 sources \cite{tevcat}. 
MAGIC has detected $\gamma$-ray emission from more than a half (39/70) of those sources (see the left panel of Fig.~\ref{fig:map}), including the two most distant ($z\sim1$) sources.  
Thanks to the excellent low energy performance of MAGIC, it is not surprising that this list contains also all but one currently known VHE $\gamma$-ray FSRQs: PKS1510-089 \cite{al14a}, 3C 279 \cite{al08a}, 3C +21.35 \cite{al11a}, B0218+357 \cite{ah16} and PKS1441+25 \cite{ah15}.
The list, status as of June 2017, of extragalactic sources detected with MAGIC is shown in Table~\ref{tab:src}. 
\begin{table}
\begin{tabular}{c|c|c|c|c|c}
Name & RA & Dec & z & period & reference \\\hline\hline
1ES0033+595 & 00 35 16.8 & +59 47 24.0 & 0.467 & 2009.08-09 & \cite{al15a} \\\hline
S20109+22 & 01 12 05.8 & +22 44 39 & >0.35 & 2015.07 & \cite{mi15} \\\hline 
RGBJ0136+391 & 01 36 32.5 & +39 06 00 & >0.27 & 2009.11 & in prep.\\\hline
B0218+357 & 02 21 05.5 & +35 56 14 & 0.954  & 2014.07 & \cite{ah16}\\\hline
3C 66A & 02 22 41.6 & +43 02 35.5 & >0.3347 & 2007.08-12 & \cite{al09} \\\hline
IC310 & 03 16 43.0 &+41 19 29 & 0.0189 & multiple & \cite{al10a} \\\hline
NGC1275 & 03 19 48.1 & +41 30 42 & 0.017559 & multiple & \cite{al12b}\\\hline
1ES0647+250 & 06 50 46.5 & +25 03 00 & $0.45\pm0.02$ & 2009-2011 & in prep. \\\hline 
S50716+714 & 07 21 53.4 & +71 20 36 & $0.31\pm0.08$ & multiple & \cite{an09a} \\\hline 
1ES0806+524 & 08 09 59 &+52 19 00 & 0.138 & 2011.01-03 & \cite{al15b}\\\hline
RBS0723 & 08 47 12.9 & +11 33 50 & 0.198  & 2014.01 & in prep. \\\hline
S40954+65 & 09 58 47.00 & +65 33 55 & >0.45 & 2015.02 & \cite{mi15b} \\\hline
1ES1011+496 & 10 15 04.1 & +49 26 01 & 0.212  & multiple & \cite{ah16a} \\\hline
Mrk421 & 11 04 19 & +38 11 41& 0.031 & multiple & \cite{ah16b} \\\hline
Mrk180 & 11 36 26.4 & +70 09 27& 0.045 & 2006.03 & \cite{al06a} \\\hline
RXJ1136.5+6737 & 11 36 30.1 & +67 37 04 & 0.1342 & 2014.01-05 & \cite{mi14b}\\\hline
1ES1215+303 & 12 17 48.5 & +30 06 06 & 0.13 & 2011.01-02 & \cite{al12a} \\\hline
1ES1218+304 & 12 21 26.3 & +30 11 29 & 0.182 & multiple & \cite{al06b} \\\hline
MS1221.8+2452 & 12 24 24.2 & +24 36 24& 0.218 & 2013.04-05 & \cite{co13} \\\hline
4C+21.35 & 12 24 54.4 & +21 22 46& 0.432 & 2010.06 & \cite{al11a} \\\hline
M87 & 12 30 47.2 & +12 23 51& 0.0044  & multiple & \cite{al08b} \\\hline
3C279 & 12 56 11.1 & -05 47 22& 0.5362  & multiple & \cite{al08a} \\\hline
PKS1424+240 & 14 27 00 & +23 47 40 & 0.604 & 2009-2011 & \cite{al14b}\\\hline
H1426+428 & 14 28 32.6 & +42 40 21 & 0.129 & 2013 & in prep. \\\hline
PKS1441+25 & 14 43 56.9	& +25 01 44& 0.940  & 2015 & \cite{ah15} \\\hline
PKS1510-089 & 15 12 52.2 & -09 06 21.6 & 0.361  & multiple & \cite{al14a} \\\hline
PG1553+113 & 15 55 44.7	& +11 11 41& 0.43-0.58  & multiple & \cite{al12c} \\\hline
Mrk501 & 16 53 52.2 & +39 45 37 & 0.034  & multiple & \cite{al07} \\\hline
H1722+119 & 17 25 04.3 & +11 52 15 & >0.35 & 2013.05 & \cite{ah16c} \\\hline
1ES1727+502 & 17 28 18.6 & +50 13 10 & 0.055 & 2011.05-06  & \cite{al14c} \\\hline
1ES1741+196 & 17 44 01.2 & +19 32 47 & 0.084 & 2010.04-2011.05 & \cite{ah17c}\\\hline
OT 081 & 17 51 32.8 & +09 39 00 & 0.322 & 2016.07 & \cite{mi16a} \\\hline
1ES1959+650 & 19 59 59.8 & +65 08 55 & 0.048  & multiple & \cite{al06c} \\\hline
MAGIC J2001+435 & 20 01 13.5 & +43 53 03 & $0.18\pm0.04$ & 2010.06 & \cite{al14e} \\\hline
1ES2037+521 & 20 39 23.5 & +52 19 50 & 0.054  & 2016.09 & \cite{mi16b} \\\hline
PKS2155-304 & 21 58 52.7 & -30 13 18 & 0.116  & 2006.07-08 & \cite{al12d}\\\hline
BL Lacertae & 22 02 43.3 & +42 16 40 & 0.069  & multiple & \cite{al07b} \\\hline
B32247+381 & 22 50 06.6	& +38 25 58 & 0.1187  & 2010.09-10 & \cite{al12e} \\\hline
1ES2344+514 & 23 47 04.9 & +51 42 17 & 0.044  & multiple & \cite{al07c} \\\hline
\end{tabular}
\caption{List of 39 sources from which MAGIC has observed $\gamma$-ray emission in years 2004-2017. Individual columns: name, right ascension and declination (taken from \cite{tevcat}), redshift, observation time period, reference.}\label{tab:src}
\end{table}

The increase of number of sources observed in the recent years in VHE $\gamma$-ray range motivated the MAGIC Collaboration to modify the observation strategy from source hunting towards deeper and more goal-oriented observations of few selected objects. 
Yearly, the MAGIC Collaboration is devoting about 40\% of the total telescopes' time (i.e. $\sim$380\,hr of dark hours and $\sim 200$\,hr of time during moonlight conditions) on observations of AGN objects. 
The MAGIC AGN program involves:
\begin{itemize}
\item long term monitoring of selected sources, among which the FSRQ PKS1510-089 (see also \cite{si17}), the HBL PG 1553+113, and the radio galaxy M87.
\item expanding the catalog of poorly populated in VHE $\gamma$-rays non-HBL AGN classes.
\item broadband multiwavelength campaigns on strong $\gamma$-ray emitters such as Mrk 421 or Mrk 501.
\item vast (over 100\,hr yearly) follow up program of mainly GeV, X-ray and optical alerts (see also \cite{gl17,za17}).
\item usage of selected AGN objects for fundamental physics studies, such us constraints on the extragalactic background light radiation (see also \cite{mo17}) or search for AGN halos due to intergalactic magnetic fields.
\end{itemize}
The MAGIC AGN observations are usually supported by GeV observations of \textit{Fermi}-LAT \cite{at09}, X-ray and UV observations of \textit{Swift} \cite{bu04} performed via ToO alerts, in-house optical monitoring and follow-up with KVA telescope \cite{ni17}. 
Often also lower frequencies (IR, radio) and optical polarization information is gathered during such studies.

\subsection{Monitoring of M87}
\begin{figure}
\centering
\includegraphics[trim = 200 180 0 0, clip=true,width=0.5\textwidth]{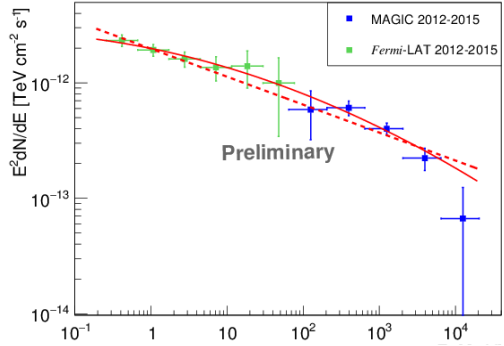}
\caption{
Spectral energy distribution of M87 measured by MAGIC (blue) and \textit{Fermi}-LAT (green). 
Dashed and solid lines show a joined spectral fit with power-law and log-parabola spectral shape respectively. 
}\label{fig:m87}
\end{figure}

M87, located at the distance of $\sim16$\,Mpc \cite{fe00} is the best studies radio galaxy known in VHE $\gamma$ rays. 
MAGIC performs a regular, deep  monitoring of M87. 
In years 2012-2015 in total over 150\,hr were gathered from this source, allowing an independent detection with above $5\sigma$ signal in each year. 
Daily and monthly light curve does not show any significant variability. 
The spectral energy distribution of the source connects smoothly with the \textit{Fermi}-LAT measurements performed in the same time period (see Fig.~\ref{fig:m87}).
The SED of the source is studied in terms of a leptonic model based on scenario presented in \cite{am11}. 
The observed multiwavelength data are consistent with a constant injection of accelerated particles at the distances of a few hundred gravitational radii from the black hole.
The spectrum of the electrons evolves as the particles propagate along the jet cooling down in SSC scenario. 
The details of this study will be discussed in a separate publication. 

\subsection{Monitoring of PG 1553+113}
\begin{figure}
\centering
\includegraphics[width=0.9\textwidth]{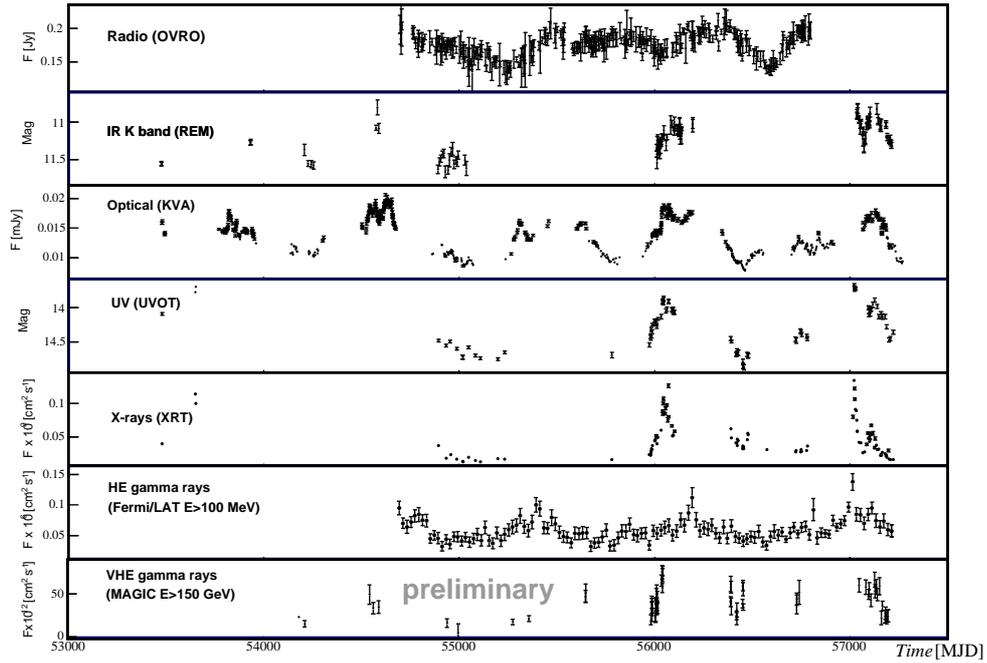}
\caption{
Multiwavelength light curve of PG1553+113. 
From top to bottom: Radio, IR, optical, UV, X-ray, GeV and sub-TeV energy ranges. 
\textit{Fermi}-LAT data are binned in 20 days bins. 
}\label{fig:1553}
\end{figure}

PG 1553+113 is a distant ($z\sim0.5$, \cite{da10}) HBL. 
Quasi-periodic oscillations with a period of $2.18\pm0.08$ year has been observed from this source in GeV, optical and partially also in radio bands \cite{ac15}.
PG 1553+113 is being observed by the MAGIC telescopes already since over 10 years (see Fig.~\ref{fig:1553}), however the early observations were irregular and biased by observations of flares. 
Regular, denser monitoring is ongoing since 2015. 
In order to explain the broadband data we consider a geometrical model of emission for periodic blazar \cite{so17}. 
According to it the variability is produced by the modulation of the observation angle due to jet precession. 
The details of this study will be discussed in a separate publication. 

\subsection{Expanding the population of LBLs and IBLs} 
Despite the relatively large population of HBLs in VHE $\gamma$ rays still only a handful of LBLs and IBLs are known at those energies. 
The small number is most probably an observational bias, as due to the shift of the  high energy peak to lower energies in those sources, they are difficult targets for IACTs. 
MAGIC, due to its excellent low energy performance can contribute to the studies of those sources and their connection to HBLs and FSRQs. 

\begin{figure}
\centering
\includegraphics[width=0.45\textwidth]{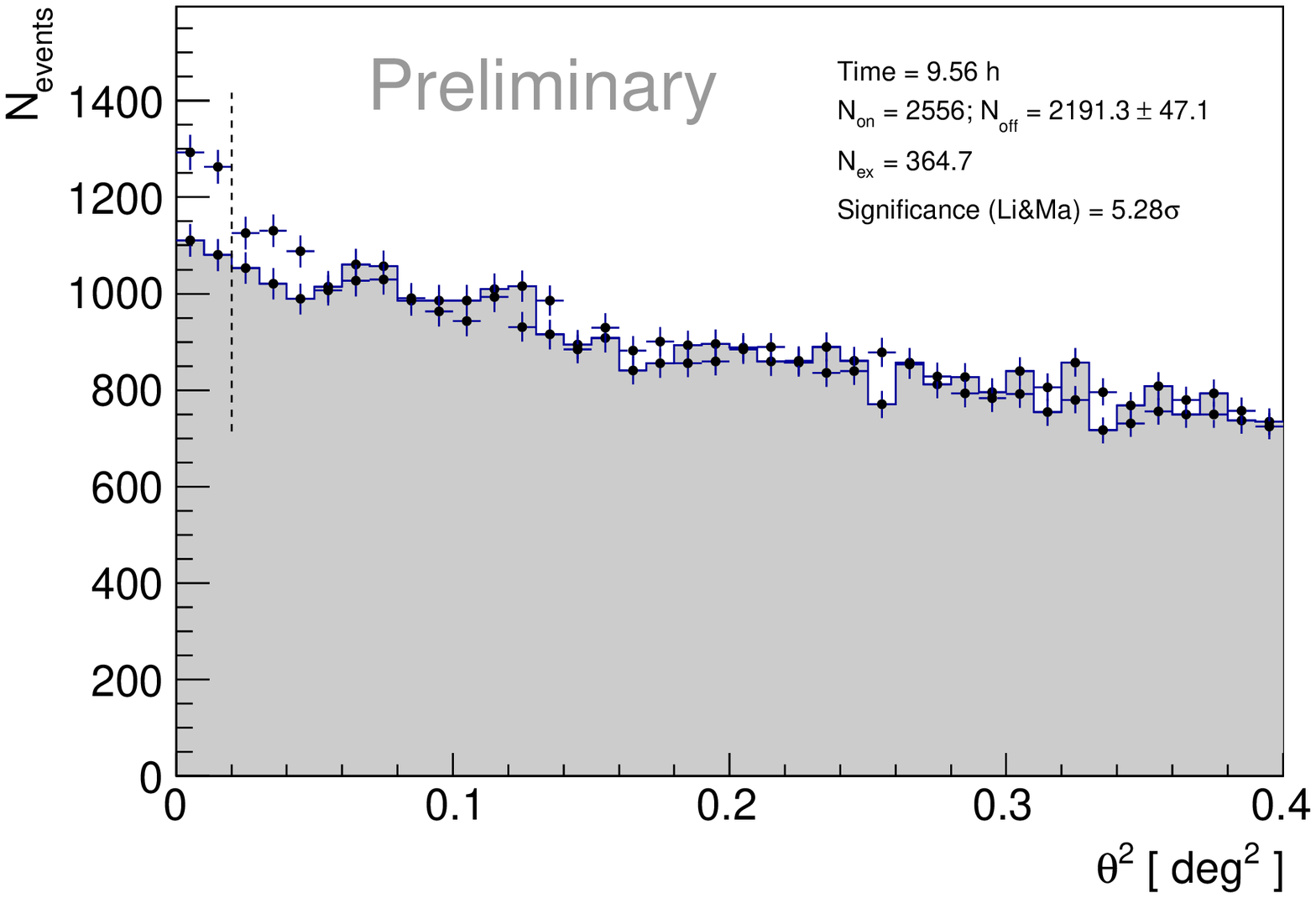}
\includegraphics[width=0.45\textwidth]{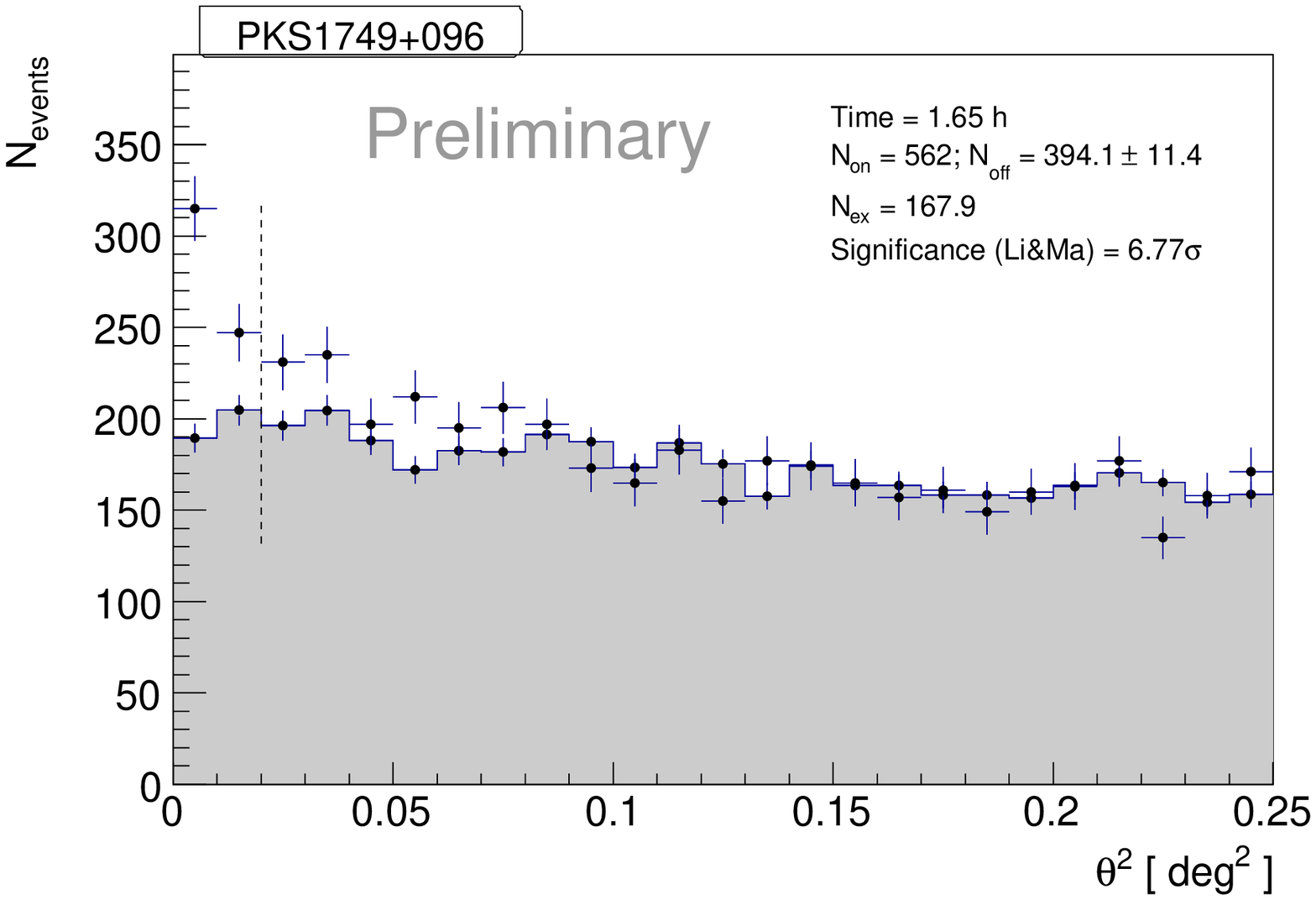}
\caption{
Distribution of the distance from the nominal source position (points) and the background estimation (shaded area). 
Left panel: S2 0109+22 observations in July 2015.
Right panel:  OT 081 observations in July 2016.
}\label{fig:det}
\end{figure}

OT 081 is a blazar classified as a LBL \cite{ac15b}.  
MAGIC observations of this source performed during elevated GeV, X-ray and optical emission revealed a significant $\gamma$ ray excess from the direction of this source \cite{mi16a}, see the right panel of Fig.~\ref{fig:det}.
S2 0109+22 is an IBL located at an unknown distance. 
The observations by MAGIC during a high state in July 2015 led to a detection of a significant $\gamma$-ray signal from this source \cite{mi15}, see the left panel of Fig.~\ref{fig:det}.

\section{Conclusions}
For the last 13 years the MAGIC telescopes have been investing a significant fraction of their observation time in studies of AGN objects.
MAGIC has detected VHE $\gamma$-ray emission from more than a half of all the currently known in this range AGN objects, for 27 objects for the first time. 
Recently the MAGIC AGN program is more focused on deep, multiwavelength studies of known VHE $\gamma$-ray emitters, and on search for new $\gamma$-ray emitters in poorly populated in this range AGN classes. 
MAGIC will continue those studies in the future preparing for the CTA era. 

\section*{Acknowledgements}
We would like to thank the IAC for the excellent working conditions at the ORM in La Palma. We acknowledge the financial support of the German BMBF, DFG and MPG, the Italian INFN and INAF, the Swiss National Fund SNF, the European ERDF, the Spanish MINECO, the Japanese JSPS and MEXT, the Croatian CSF, and the Polish Narodowe Centrum Nauki.

\def\aj{AJ}
\def\aap{A\&A}
\def\apj{ApJ}
\def\apjl{ApJL}
\def\procspie{Proc. SPIE}
\def\mnras{MNRAS}
\def\aaps{A\&AS}

\end{document}